\documentclass[12pt,preprint]{aastex}

\slugcomment{ARTIST, 09 Mar 2012}

\begin{document}

\title{ADAPTIVE REAL TIME IMAGING SYNTHESIS TELESCOPES}

\medskip

\author{Melvyn Wright}
\affil{Radio Astronomy Laboratory, University of California, Berkeley,
    CA, 94720}

\begin{abstract}
The digital revolution is transforming astronomy from a data-starved
to a data-submerged science.  Instruments such as the Atacama Large
Millimeter Array (ALMA), the Large Synoptic Survey Telescope (LSST),
and the Square Kilometer Array (SKA) will measure their accumulated data
in petabytes. The capacity to produce enormous volumes of data must
be matched with the computing power to process that data and produce
meaningful results. In addition to handling huge data rates, we need adaptive
calibration and beamforming to handle atmospheric fluctuations and radio
frequency interference, and to provide a user environment which makes
the full power of large telescope arrays accessible to both expert and
non-expert users.  Delayed calibration and analysis limit the science
which can be done.  To make the best use of both telescope and human
resources we must reduce the burden of data reduction.

We propose to build a heterogeneous computing platform for real-time
processing  of radio telescope array data.  Our instrumentation comprises
of a flexible correlator, beam former and imager that is based on
state-of-the-art digital signal processing closely coupled with a
computing cluster.  This instrumentation will be highly accessible
to scientists, engineers, and students for research and development of
real-time processing algorithms, and will tap into the pool of talented and
innovative students and visiting scientists from engineering, computing,
and astronomy backgrounds.  The instrument can be deployed on several
telescopes to get  feedback  from dealing  with real sky data on working
telescopes.

Adaptive real-time imaging  will transform radio astronomy by providing
real-time feedback to observers.  Calibration of the data is made in close
to real time using a model of the sky brightness distribution. The
derived calibration parameters are fed back into the imagers and
beam formers. The regions imaged are used to update and improve the
a-priori model, which becomes the final calibrated image by the time
the observations are complete.

\end{abstract}
 
\section{INTRODUCTION}

The scientific goals of  the next generation of radio telescopes will be
enabled by transforming our approach to signal processing by exploiting
the digital revolution.  Real-time signal processing for telescope arrays
must address data rates that will exceed $\sim$1 terabyte/s and require
petaop/s signal processing.  The huge volumes of data must be matched
with the computing power to process that data and produce meaningful
results.  Innovative approaches in signal processing, computing hardware,
algorithms, and data handling are necessary. In addition to handling
the data rates, adaptive calibration and beamforming are essential to
handle atmospheric perturbations (adaptive optics), and radio frequency
interference (RFI), and to provide a user environment which makes the
full power of large telescope arrays accessible to both expert and
non-expert users.

The current data processing paradigm uses on-line custom digital signal
processing (DSP) with off-line data reduction and analysis in general
purpose computers.  Off-line  processing can handle only a few percent
of the data generated by the on-line DSP.  The large time between data
acquisition and analysis, results in lost science opportunities.

In this paper we propose to address these problems  by integrating
on-line and off-line data processing in a heterogeneous system using
ASIC, FPGA, GPU  and computer clusters to provide a  flexible programming
environment with real-time feedback. Adaptive real-time imaging
enables us to image large regions with high frequency and time resolution.
Variable sources, instrumental problems and RFI are handled in real time.
We propose to build a development computing platform with a flexible
correlator, beam former and imager for radio telescope and receiver
arrays that is based on state-of-the-art, digital signal processing
closely coupled with a computing cluster.  This instrumentation will
be accessible to scientists, engineers, and students for research
and development of real-time processing algorithms, and taps into the
pool of talented and innovative students and visiting scientists from
engineering, computing, and astronomy backgrounds.  Adaptive real-time 
imaging  is a major step in transforming synthesis imaging from an off-line to 
a real-time process --- a digital camera for radio telescopes. This transformation
enables new science, and  is necessary to prevent astronomers from being
overwhelmed by data and off-line data reduction.  In addition to signal
processing and scientific advances, new approaches are needed to enable
power-efficient instrumentation that is affordable on a massive scale.
Adaptive real-time imaging will revolutionize the science capabilities of existing
and developing telescopes like the Atacama Large Millimeter Array (ALMA),
the Murchison Wide-Field Array (MWA), and have a broad impact on the
way that radio telescope arrays can be used.  Adaptive real-time imaging  
will transform synthesis telescopes by providing real-time feedback to observers.
Obtaining calibrated data and images quickly  will enable astronomers to optimize
the observations and calibrations needed to realize their science.

Section 2 reviews the current state of the art and the problems faced
by existing and next generation aperture synthesis telescopes. Section 3
presents a model for developing adaptive real-time imaging. In Section 4,
5 and 6 we trace the data processing from the telescopes through cross
correlation, calibration and imaging. Section 7 presents some current
developments and conclusion.

\section{APERTURE SYNTHESIS IMAGING}

Arrays of radio telescopes enable us to map the sky brightness using
aperture synthesis techniques (Thompson, Moran, \& Swenson, 2001
[TMS2001]).  If the dimensions of the radio source and the telescope
array are small compared with the distance to the source, then the
coherence of the wavefront is proportional to the Fourier transform of the
intensity distribution of the source (Van Cittert-Zernike theorem, Born \&
Wolf, 1959). The coherence, also known as the visibility  function, is
obtained from measurements of the cross correlation of signals between
pairs of antennas in the telescope array. 
A telescope array with $N$ antennas, provides $N(N-1)/2$ cross
correlations and $N$ auto correlations for each polarization product.
The Earth's rotation of  the projected geometry of the telescope array
in the direction of a celestial source provides additional samples of the
source visibility function in the aperture plane (Ryle, 1962; TMS2001).

Digital cross correlators  compute cross-power frequency spectra for
all pairs of antennas in the telescope array.  Since the signals from
celestial radio sources are typically much weaker than the uncorrelated
noise power from sky and radio receivers, the measured cross correlations
are time-averaged to enhance the signal-to-noise ratio.

The Expanded VLA (EVLA)  with 27 antennas (Perley et al. 2011), and
the Atacama Large Millimeter Array (ALMA)  with 64 antennas (Wootten
\& Thompson, 2009) represent  the current state of the art aperture
synthesis telescopes at centimeter and millimeter/submillimeter
wavelengths respectively.

The digital correlators are peta-op, special-purpose computers. The
EVLA correlator (Carlson \& Dewdney, 2000) cross correlates all pairs of
antennas with up to 16 GHz of bandwidth with a minimum of 16,384 spectral
channels in 64 full polarization, independent spectral windows. The ALMA
correlator (Escoffier et al., 2007) processes 16 GHz of bandwidth for the
2016 pairs of antennas and 4 polarization products. The basic operation is
a complex-multiply and add operation. The complex multiply is typically 
4$\times$4-bit with accumulation into 32-bits at rates $\sim10^{17} $s$^{-1}$.
Large  digital correlators built using custom ASICS take 5-10 years to
develop (e.g., ALMA: Escoffier et al. 2007; EVLA: Perley et al. 2011).

Time-averaged correlation data  are written to a data archive for
off-line data processing.  The data rate from the EVLA correlator can
be up to 350 GB s$^{-1}$.  Only a few percent of this data rate can be
handled by the off-line data processing. The current plan for  ALMA is an 
average data rate $\sim$ 6 MB  s$^{-1}$  and a peak rate  60 MB s$^{-1}$ 
(Lucas et al. 2004). Even so, users will be faced
with the prospect of dealing with several terabytes of data for EVLA
and ALMA observations (EVLA: Perley, 2004; ALMA: Lucas et al. 2004).

Calibration and imaging are made in general purpose floating point
processors using the averaged cross correlations from the data archive.
The complex-valued cross correlations are samples of the Fourier
transform of the sky brightness distribution.  These are calibrated
w.r.t. measurements of known sources.  When sufficient cross-correlations
have been measured, images of the sky brightness distribution can be
made from the Fourier transform of the calibrated cross correlations. The
Images, $I({\bf s}, f, p, t)$,  are, in general, functions of position,
${\bf s}$, frequency, $f$, polarization, $p$  (Stokes I, Q, U, V),
and time, $t$.

Sophisticated image processing algorithms have been developed to
self-calibrate the measured cross-correlation function using images
of the sky brightness, and to remove sidelobes of the synthesized
beam and confusing sources (e.g., TMS2001; Cornwell and Perley, 1992).
These algorithms have been very successful, but are time consuming and
require a level of expertize which many astronomers do not wish to acquire
in order to do their science.  The delayed calibration and analysis
of the data limit the science which can be done.  Variable sources,
targets of opportunity, instrumental and atmospheric problems, and
radio frequency interference (RFI) at low frequency, are more easily 
handled as the data are being acquired.

Aperture synthesis arrays at meter wavelengths present formidable
problems.  The wide field of view of the telescopes are full of
radio sources which confuse the regions of interest. The antennas have
direction-dependent response over the field of view, and the ionosphere
can cause direction-dependent phase shifts on short time scales. LOFAR
is a Low Frequency Array telescope with antennas at 77  stations spread
over 100 km and observing in the frequency range 30-90 and 120-250
MHz. Data from the antennas at each station are combined into phased
array beams to reduce the data rate to a single data stream for each
station. Correlation of the station beams is made in a 34 TFlop, IBM
BlueGene/L. LOFAR calibration and imaging are made in pipelined data
processing performing RFI flagging, with calibration using a model sky
brightness  model (Nijboer \& Noordam, 2006). The Murchison Wide-Field
Array (MWA) was designed as a 512-antenna array being built in Western
Australia to observe in the frequency range 80-300 MHz. The correlation
data would comprise of 130,000 cross correlations with 768 frequency
channels and 4 polarization products (Ord et al. 2009). The data
rate $\sim$19 GB s$^{-1}$ is impractical to store;  data will be calibrated and
imaged in a real-time pipeline with images of the sky produced at 8 s
intervals. The real-time calibration pipeline processing is discussed
in detail by Mitchell et al. (2008). The MWA was recently de-scoped to
a 128-antenna array, which reduces the data rate by a factor 16.

 The Square Kilometer Array (SKA) will be able to form simultaneous
images in multiple regions within the field of view.  The SKA science
requirements (Schilizzi et al. 2007) require imaging multiple regions
with an image fidelity  $\sim10^4$ between 0.5 and 25 GHz.  The bandwidth
$\sim$25\% at observing frequencies below 16 GHz, and 4 GHz above 16
GHz. Each band will have $\sim10^5$ spectral channels with a minimum
accumulation interval 0.5 s.  The images should have at least $10^5$ beam
areas at the maximum angular resolution.  Three Key Science projects
require all-sky  surveys.  Survey science requires images with superb
image quality, which imposes stringent requirements on the calibration
and sidelobe levels at every stage of beam formation.  A major theme
driving the SKA design is the high cost of data processing (Perley et
al. 2003; Cornwell 2004, 2005;  Lonsdale et al. 2004; Wright et al. 2006).

\section{DATA PROCESSING MODEL }

In this paper we propose to develop calibration and imaging in close
to real time in order to reduce the burden of expert data reduction on
the end user, and to make best use of both telescope and human resources.
Large arrays and new science require seamlessly integrating calibration
and imaging into the data acquisition process.  Calibration, imaging,
and deconvolving the response of sources outside the fields of interest
are intimately related, and are best handled in close to real time,
rather than using off-line data processing.  Calibration in close to
real time uses a model of the sky brightness distribution.  The derived
calibration parameters are fed back into the imagers and beam formers. The
regions imaged update and improve the a-priori model, which becomes the
final calibrated image by the time the observations are complete (Figure 1).

High performance digital signal processing enables us to handle high data
rates in parallel, and to make images in close to real time. Images can
be made simultaneously for multiple regions within the field of view
by integrating the output from the correlators on multiple targets of
interest, calibration sources, and sources whose sidelobes confuse the
regions of interest.

The system design uses modular DSP boards with a 10 GbE interconnect
architecture which allows reconfiguration of the computing resources for
multiple applications.  The programming model uses a system generation
library with hardware abstractions which allow the application programmer
to focus on the application rather than the details of the hardware.
The system design and programming model together allow the application
software to survive by using a technology independent design flow.
A major problem in the design of large antenna array processors has
been the routing of high-bandwidth data. Each cross-correlation
processor and beam former must receive data from every antenna,
and the number of interconnections can become unmanageable. The
CASPER\footnote{Collaboration for Astronomical Signal Processing
and Engineering Research; http://casper.berkeley.edu} group has
developed a packetized signal flow architecture capable of performing
this antenna/frequency data transposition using commercial 10 Gbit
Ethernet (10 GbE) switches (Parsons et al. 2008). The FPGA devices are
programmed using open-source signal processing libraries developed and
supported at multiple observatories that allow flexible, scalable, and
device-independent solutions (Brodersen et al. 2004; Parsons et al. 2008).
This ongoing work reduces the time and cost of implementing interferometer
processors while supporting upgrades to new generations of processing
technology.

A hybrid solution using beam formation and correlators provides a flexible
development path for imaging large fields of view. Phased array beams can
be formed anywhere in the sky by adding the signals from the antennas.
The sidelobe structure of each beam depends on the array geometry, the
source direction, and the amplitude and phase weighting of the signals
from each antenna. Beam formation is appropriate for analyzing signals
from discrete radio sources such as pulsars, SETI targets and RFI sources.
Beam formation allows us to channel the collecting area of large arrays
of telescopes into expensive back-end analysis engines.  Direct imaging
using beam formation is appropriate for compact sources, but is currently
too expensive for imaging large fields.  Correlators provide a versatile
mechanism for imaging multiple regions within a field of view.

\section{DATA FLOW }

In this section we trace the data flow through an imaging system using
correlators and beam formers.  Figure 1 shows the overall system.
The total bandwidth of signals from $N$ antennas is $N \times B \times
N_{pol}$, where $B$ is the analog bandwidth and $N_{pol}$ the number
of polarizations from each antenna.
The data for each antenna are digitized with 2-12 bit precision.
The total data bandwidth is $N \times 2 B \times N_{pol} \times N_{bits}$,
e.g., for $N$ = 1000, $B$ = 1 GHz, $N_{pol}$ = 2, and $N_{bits}$ = 8,
the total data bandwidth is $4~10^{12}$  bytes s$^{-1}$.  

The bandwidth must be channelized, to provide spectral resolution, to
facilitate interference rejection, and to reduce bandwidth smearing by
using multi-frequency synthesis (MFS).  The science and RFI requirements
for a large number of frequency channels favor an `FX' architecture
(Figure 1).  Voltage signals from each antenna and polarization are divided
into many frequency channels (`F' stage).  Excellent separation of
frequency channels can be obtained with polyphase filters.  Fixed point
processors are well matched to the `F' stage, with $\sim log(N_{chan})$
complex-multiply-add operations and a high data bandwidth.  After the
frequency transform, the data can be processed in parallel, reducing
the data rate in each frequency channel by a factor $N_{chan}$.

These data are routed into cross correlators for each pair of antennas
and frequency channel to measure the correlation properties of the
incident radiation (`X' stage), and into beam formers to form phased
array beams at multiple points in the sky.  Commercial 10-GbE  switches
provide flexible routing which allows the DSP to be upgraded, repaired,
and reprogrammed with minimum interruption to telescope operations.

Cross correlation is a  complex-multiply and accumulate operation for
all pairs of antennas and polarizations.  For a dual polarization
array with $N$ antennas and bandwidth, $B$, the correlator must
provide $2N(2N+1)/2 \times B$ complex-multiply and accumulate
operations per second, independent of the number of frequency channels.
The complex-multiply is typically  4$\times$4-bit with accumulation into
32-bits. Using fewer bits in the cross correlation results in a small
loss in resulting signal-to-noise, but allows the use of lookup tables
for the cross correlations. Floating point processors can also be used.
Performance, cost and power comparisons of ASIC, FPGA, GPU, and CPU
processors have been made by a  number of authors (Ord et al. 2009;
Nieuwpoort \&  Romein 2009; Clark, LaPlant, \& Greenhill 2011).
An order of magnitude estimate for the cost of a custom
ASIC correlator which has been considered by the CASPER group,
is \$2M + \$10 per chip, versus \$2000 for a high performance GPU or FPGA,
which suggest that an ASIC correlator would be a better solution for
systems with more than $\sim$ 1000 chips. The development times are:
GPU ($\sim$ 1 yr), FPGA ($\sim$ 2 yr), ASIC ($\sim$ 5 yr).  
ASICs or FPGAs offer more bandwidth, GPUs offer more FLOPS, per \$. 
Development time favors GPUs. ASICs will be required for arrays with large
numbers of correlations in order to meet the power/heat requirements.
A heterogeneous system
would allow choosing the appropriate solutions at each stage of the
data flow. Clark et al. (2011) have a useful discussion in section 2.3.

After correlation there are $N(N+1)/2$ auto and crosscorrelations for
each polarization product. The data are then time averaged.  The data
rate is reduced from the input bandwidth, $B$ to the rate of change of
the cross correlations -- the fringe rate.  In order to correlate the
signals from a siderial source anywhere in the sky, the data bandwidth
from the correlator is:

$N(N+1)/2 \times N_{pol} \times N_{chan} \times N_{bits} \times 2~sdot \times D_{max}/\lambda $,

where  $N_{pol} = 4$ polarization products, 
and $sdot$ is the earth rotation rate, $7.27~10^{-5}$ radian s$^{-1}$.
e.g., for $N = 1000$, $N_{chan} = 10^5$, $N_{pol} = 4$, $D_{max} =
1000~km$, $\lambda = 1~cm$, and $N_{bits} = 2 \times 16$ (complex data),
the total data bandwidth would be $\sim 10^{16}$ bytes s$^{-1}$.

Sampling the correlator at the fringe rate allows us to make images
over a wide field of view, including targets of interest, calibration
sources, and sources whose sidelobes confuse the regions of interest.
We can form simultaneous images in multiple regions within the field of
view by integrating the output from the correlators at multiple phase
centers. The data stream from each correlator is multiplied by phase
factors, $\exp(2\pi i /\lambda ~r.s_o)$, where $r$ = $(r_j - r_k)$
is the baseline vector for antenna pair $(r_j , r_k)$, and $s_o$ is
the phase center in each region of interest.  The data bandwidth for
imaging the primary beam width is:

$N(N-1)/2 \times N_{pol} \times N_{chan} \times N_{bits} \times 2~sdot
\times D_{max}/D_{ant}$.

e.g., for $N = 1000$, $N_{chan} = 10^5$, $N_{pol} = 4$, $D_{ant} = 12~m$,
and $N_{bits} = 2 \times 16$ (complex data), the total data bandwidth
is $2~10^{10}$ bytes s$^{-1}$ for 1 km baselines, and $2~10^{13}$ bytes
s$^{-1}$ for 1000 km baselines.

\section{CALIBRATION}

An a-priori model of the sky brightness distribution is used for
calibration and imaging. In the standard observing paradigm, strong
compact sources are used as primary calibrators, and self-calibration
is used to improve the calibration during the off-line imaging process.
The calibrations are the product of antenna station beam patterns, gains,
bandpass and polarization corrections which are derived from a least
squares fit of the data to a model visibility which is computed for the
calibration source or the sky brightness model. For compact sources,
a simple direct Fourier transform can be used. For more complex sky
brightness models, a gridded FFT can be used to derive the model
visibility used in a least squares fit to the measured cross correlations.
The calibrations may vary with time and position in the sky. For phased
array station beams, atmospheric fluctuations make the primary beam
response time variable.  Our approach to these problems is to separately
calibrate the data for each phase center.  We can identify regions
which have bright emission from a-priori images of the sky brightness,
and image only regions which are of interest or contain sources whose
sidelobes corrupt the regions of interest.  Confusing sources may be in
the sidelobes of the primary beam, or in different isoplanatic regions.
The sky model is used in a self calibration algorithm to determine the
antenna calibration as a function of time for each phase center which
contains suitable sources. The calibrations at each phase center are
correlated, and can be improved by developing a global model  across the
array as a function of time and frequency (Nijboer et al. 2006).  If the
source contains spectral lines, multiple frequency channels are used
simultaneously to determine the calibration.  Observations in multiple
frequency bands can be used to separate the gains into tropospheric and
ionospheric delays. The data streams can be buffered so that the gains
can be averaged and interpolated before being applied to the data stream.
Confusing sources are removed by subtracting the source model from the
calibrated data stream. The subtraction can be made for each region of
interest and frequency channel in distributed processors associated with
each correlation engine, but including the response from the whole sky
model, especially of course the strong sources (Wright 2005; Mitchell
et al. 2008).

The basic calibration computation is a complex-multiply of the measured
cross correlations ($uv$ data) for each data sample and frequency channel.
The calibrations can be stored in data structures and applied when
the $uv$ data are plotted, analyzed, or imaged.  In Figure 2, we plot
the computation time for calibrating multi-channel $uv$ data versus the
number of $uv$ data samples in an off-line simulation for ALMA data with
60 antennas in a 4 km configuration. Figure 2 shows that the off-line
calibration time is proportional to the number of $uv$ data samples. We
used the MIRIAD data reduction package (Sault, Teuben \& Wright, 1995),
which uses a streaming data format. The complex-valued $uv$ data were
represented by 4 bytes per frequency channel with a scaling factor
for each multi-channel data sample. The 4-byte representation of the
$N_{chan}$ allows a 1:32,000 spectral dynamic range for each multi-channel
data sample.  Including the time-variable meta-data which describe the
data, the telescope, and the observations, the total length was 460
bytes for a 100-channel data sample. The calibration rate was 6 Mbytes
s$^{-1}$, showing that the average data rate currently allowed for ALMA
could be calibrated in a single pipelined process on a standard rack
server, and that much higher data rates could be supported in multiple
threads on a modest sized cluster. Further gains in computing efficiency
are clearly possible. Off-line data reduction typically uses static
``measurement sets" with the $uv$ data represented as 8- or 16-byte
complex values. An astronomer using off-line data processsing typically
keeps several copies of calibrated and uncalibrated $uv$ data, with each
step requiring reading and writing the $uv$ data.

In a real-time imaging pipeline, the calibrations are derived from,
and applied to the data streams from the correlators (see Figure 1).
RFI must be also subtracted from the data stream before it is passed
to the imaging engine and beam formers.  RFI presents a special case
in several ways. RFI sources may be stationary, or moving across the
sky at a non-siderial rate. A correlator can be used to locate and
characterize RFI as a function of time, frequency and polarization.
The signal-to-noise can be improved by pointing some of the antennas
or beam formers at the RFI sources.  Correlators allocated to measuring
RFI may need to sample the signal at high data rates. 
For phased array telescopes, the station beam can form nulls at
the position of (moving) RFI sources. Accurate calibration of the
array antennas is required in real time (Barott et al. 2011).

\section{REAL TIME IMAGING}

The standard imaging algorithm is an FFT of the gridded $uv$ data for each
field of view, polarization and frequency channel. Here we review the
basic math.  For a more detailed description, see e.g., Thompson, Moran,
\& Swenson, (2001), and references therein.  The brightness distribution
is the Fourier transform of the sampled visibility data, $V$.  Since we
only have discrete samples of $V$,  we define a weighting function
$W$, and make an image, $I'$ which is the Fourier transform of the
product of $V$ and $W$.
The weighting function $W$ is typically chosen to minimize the noise
and make a  more uniform weighting of the sampled $uv$ plane.  $W=0$
where $V$ is not sampled. The image $I'$, the  Fourier transform of the
product of $V$ and $W$, is the convolution of the Fourier transforms of
$V$ and $W$. The observed brightness distribution is the sky brightness
distribution, $I({\bf s}, f, p, t)$, illuminated by the primary beam
pattern, $A(s, \nu ,p)$.  Omitting the functional dependence for clarity,
$I'=[I \times A] \star B$, where
$B$ is the synthesized beam, the instrumental point-source response.

In order to use a fast Fourier transform algorithm,
we re-sample the $uv$ data onto a gridded $uv$ plane. 
The $uv$ data are multiplied by the weighting function $W$,
convolved by a gridding function $C$, and re-sampled onto a regular grid by $\Pi$.
\[[(V\times W) \star C]\times \Pi  <= FFT => [((I \times A)\star B) \times c] \star \amalg\]
Thus, the Fourier transform of the gridded $uv$ data is an image of
the sky brightness distribution $I$, multiplied by the primary beam
pattern, $A$, convolved the synthesized beam $B$, multiplied by $c$,
and convolved by $\amalg$.  The convolution by $\amalg$ replicates
the image at intervals $1/\delta uv$, where $\delta uv$ is the sample
interval of the gridded $uv$ data. Aliasing in the sky brightness image
is minimized by choosing a function $C$, so that its Fourier transform
$c$ falls to a small value at the edge of the image.

The imaging step is usually followed by correction in the image plane
for the gridding convolution, $c$, and deconvolution to remove the
response to source structure in the sidelobes of the synthesized beam, $B$.
Two different deconvolution algorithms are commonly used: an iterative
point source subtraction algorithm, CLEAN, which is well matched for
deconvolving compact source structures, and MAXIMUM ENTROPY, a gradient
search algorithm, which maximizes the fit to an a-priori image, in a
least squares fit the the $uv$ data. Both algorithms operate in the
image plane on the synthesized image and beam.

In Figure 2, we plot the time for a gridded FFT in MIRIAD for a
multi-channel image with 1280 $\times$ 1280 pixels and 100 frequency
channels. The multi-channel data are gridded and imaged as a vector with
a common pixel size, gridding convolution function, and synthesized
beam.  Figure 2 shows that the imaging time is proportional to the number
of $uv$ data samples, at a rate 3.4 Mbytes s$^{-1}$ using a single
processor. Image deconvolution time, scales with image size and complexity,
and is cpu intensive. A direct deconvolution, dividing by the Fourier
transform of the synthesized beam, can not be used because the Fourier
plane is not completely sampled. Both CLEAN and MAXIMUM ENTROPY are
iterative algorithms, using FFTs of the image and synthesized beam.
A 1280 $\times$ 1280 $\times$ 100 channel, real valued image
(4-bytes per pixel) $\sim$650 Mbytes, with a common synthesized beam
(6.5 Mbyes), can be deconvolved in memory.
The frequency channels can be deconvolved in parallel processes. 
Image deconvolution is relatively fast for compact image
structures, but can exceed the imaging time for complex images.

In a real-time imaging pipeline, images in multiple frequency channels
can be processed in parallel in a distributed architecture. The images
are formed from the calibrated data stream from which the a-priori sky
model has been subtracted, and are therefore difference images from
the sky model.
 Subtracting the sky brightness model from the $uv$ data
minimizes many of the problems in the gridded FFT, and in particular
allows position dependent calibrations and time variable primary beam
patterns to be handled (see, e.g., Wright \& Corder 2008). 
The difference images are used to update the sky model, including not
only the regions of interest, but also improving the accuracy of sources
whose sidelobes must be subtracted.  As the observations proceed, both
the model image and the calibration are improved. The process converges
when the difference images approach the noise level and the model image
is consistent with the data.
For a small field of view a 2D FFT can be used to image the
region around each phase center.  The maximum image size for a 2D FFT
scales as $D_{max}/\lambda$, $\sim 10^8$ beam areas on a 1000 km baseline
at $\lambda$ 1 cm.  Deconvolution is minimized by obtaining good $uv$
sampling of the aperture plane, and low synthesized beam sidelobe levels
for large $N$ array designs. e.g., for the ALMA array with 60 antennas,
the sidelobe levels are $\sim$ 1 \%.  In many cases, deconvolution in
the image plane may not be needed, since the model image and sidelobes of
confusing sources have been subtracted from the $uv$ data. In addition,
images may be limited by atmospheric and instrumental errors which must
be removed from the $uv$ data and can not be removed by deconvolving in
the image plane.

The imaging engine can make images using all the frequency channels.
Spectral line images can be made for multiple frequency channels,
averaged into the desired frequency or velocity intervals.  Wideband,
MFS imaging treats the frequency channels as independent $uv$ samples.
The a-priori model used in the calibration can be updated at intervals,
when the difference from the best current image is significant.

Variable sources are detected as intermittent sources which are
inconsistent with the current model.  We should also accumulate a $\chi^2$
image to help identify pixels where there are time variable sources or
RFI sources. In some cases we may want to keep a time series of difference
images and the model images used for the calibration.

We view imaging as a dynamic process which can be guided in real time by
observers inspecting the convergence of the model image and the $\chi^2$
image.  As the observations proceed, the observations can be moved to
regions where more data are needed to define the science goals, either
regions of interest, or sources whose sidelobes are confusing, or new
sources which are discovered in the imaging process.  Isoplanatic patches
may vary during the observations requiring different observation centers
to adequately determine the calibration across the sky.

The data archive serves as the data base for the observations,
calibrations and instrument status during the observations.  The data
streams from each phase center are saved in the data archive along with
the metadata.  Data from the data archive can be replayed though the
imaging system so that the best model of the sky and calibration data
from the completed observations can be used to improve the calibration
of the final image and extract time variable sources.

\section{CURRENT DEVELOPMENTS}

In this paper we propose to develop adaptive real-time imaging using
correlators and beam formers with a high data bandwidth into computer
clusters.  The mismatch between the data rates in the on-line DSP and
those supported by off-line processing is resolved by integrating the
calibration and imaging with the data acquisition process.  Calibration
and imaging are handled with the real-time feedback of the antenna
calibration needed for beam formers and RFI suppression.

Images can be made simultaneously for multiple regions within the field of
view by integrating the output from the correlators on multiple targets
of interest, calibration sources, and sources whose sidelobes confuse the
regions of interest.  The regions imaged are used to update and improve
the a-priori model, which becomes the final calibrated image by the time
the observations are complete.

A number of current telescopes are developing these concepts.
The Allen Telescope Array (Welch et al. 2009) is a leading prototype
for the SKA. Small (6.1 m) dishes give the ATA excellent survey speed
for wide field imaging, with a  frequency coverage from 0.5 to 11.2
GHz.  The use of flexible digital signal processing enables multiple
simultaneous observing projects and automated data processing (Keating
et al. 2010). Figure 3 shows an example of the data processing required
for transient sources.

The Precision Array for Probing the Epoch of Re-ionization (PAPER) is an
array of precision dipoles to map the whole sky  which uses a packetized
correlator design (Parsons et al. 2008). Calibration uses an all-sky model
(Parsons et al. 2010).

The Murchison Wide-Field Array (MWA) is a low-frequency radio telescope
to search for the spectral signature of the epoch of reionization (EOR)
and to probe the structure of the solar corona. The MWA will have 128
antenna arrays capable of imaging the sky from 80 MHz to 300 MHz with an
instantaneous field of view that is tens of degrees wide and a resolution
of a few arcminutes (Mitchell et al. 2008). A data rate $\sim$1 GB/s  with
images every  8 s requires on-site, real-time processing and reduction
in preference to archiving, transport and off-line processing. Real
time performance needs $\sim$2.5 TFLOP/s. Edgar et al (2010) present a
heterogeneous computing pipeline implementation, using  GPUs which are a
good fit for pipeline processing, but lack flexibility or feedback into
the data acquisition, e.g., RFI detection and excision.

High performance digital signal processing enables us to handle high data
rates  from aperture synthesis arrays in parallel,  and to make images in
close to real time. Adaptive real-time data processing will revolutionize
the science capabilities of existing and developing telescopes, and have
a broad impact on the way that radio telescope arrays can be used.

The current situation for aperture synthesis arrays may be compared with a
jet-liner. Both are extremely complicated and sophisticated systems. Both
can be programmed to function automatically to deliver the expected
results. However, the pilot of the jet-liner has the full power of the
control systems to handle unexpected situations in real time. Whereas,
the astronomer has little real-time feedback, or ability to adapt the
observations to new discoveries, and, even worse, the unexpected result
may not be in the data because off-line data processing can not handle
the data rate or computational needs.  High performance computing
with real-time feedback to observers  will enable us to optimize the
observations and calibrations needed to realize the science.

\section*{Acknowledgements}

I thank the many colleagues and students who have contributed to this
work, in particular, Dan Werthimer and the CASPER students and staff
for hardware and software tools which enable us to build reconfigurable
data processing for radio astronomy. Thanks to the anonymous referees
for valuable comments which have improved the presentation of this
work. This paper is dedicated to my colleague, the late Don Backer.
This work was supported by the National Science Foundation under grant
number AST-0906040, ``Collaborative Digital Instrumentation for the
Radio Astronomy Community"

\clearpage

\section{ References }

Barott, W. C., et al., 2011, Radio Science, 46, 1016
``Real-time beamforming using high-speed FPGAs at the Allen Telescope Array"

Born, M., \& Wolf, E.,\ 1959, London: Pergamon Press

Brodersen, B., Chang, C., Wawrznek, J., Werthimer, D., \& Wright, M., 2004,
``BEE2: A Multi-Purpose Computing Platform for Radio Telescope Signal Processing Applications"
http://bwrc.eecs.berkeley.edu/Research/BEE/BEE2/presentations/BEE2\_ska2004\_poster.pdf

Carlson, B. R., \& Dewdney, P. E., 2000, Electron lett., 36 987

Clark, M. A., LaPlante, P. C., \& Greenhill, L. J., 2012, ``Accelerating Radio Astronomy Cross-Correlation with Graphics Processing Units," International Journal of High Performance Computing Applications,   arXiv:1107.4264

Cornwell, T.J. \& Perley, R.A., 1992, ``Radio-Interferometric Imaging of Very Large Fields",
A\&A 261, 353

Cornwell, T.J., 2004, EVLA memo 77
``EVLA and SKA computing costs for wide field imaging (Revised)"

Cornwell, T.J., 2005, SKA memo 64, http://www.skatelescope.org
``SKA computing costs for a generic telescope model"

Edgar, R.~G., Clark, M.~A., Dale, K., Mitchell, D.~A., Ord, S.~M., Wayth, R.~B., Pfister, H., 
\& Greenhill, L.~J.\ 2010, Computer Physics Communications, 181, 1707

Escoffier, R.~P., Comoretto, G., Webber, J.~C., et al.,\ 2007, \aap, 462, 801


Keating, G.~K., Barott, W.~C., \& Wright, M.,\ 2010, \procspie, 7740

Lonsdale, C.J., Doeleman, S.S., \& Oberoi, D., 2004, SKA memo 54, http://www.skatelescope.org
``Imaging Strategies and Post processing Computing Costs for Large-N SKA Designs"

Lucas, R. et al., 2004, ALMA memo 501

Mitchell, D.~A.,  Greenhill, L.~J., Wayth, R.~B., et al.,\ 2008, IEEE Journal of Selected 
Topics in Signal Processing, 2, 707


 Nieuwpoort, R. V. V., \&  Romein, J. W., 2009, 
ICS 2009 Proceedings of the 23rd international conference on Supercomputing

Nijboer, R.~J.,  Noordam, J.~E., \& Yatawatta, S.~B.\ 2006, Astronomical Data Analysis Software and Systems XV, 351, 291


Ord, S., Greenhill, L.,  Wayth, R., et al.,\ 2009, Astronomical Data Analysis Software and Systems 
XVIII, 411, 127

Parsons, A., et al.\  2008, \pasp, 120, 1207.

Parsons, A.~R., et al.\  2010, \aj, 139, 1468

Perley, R., \& Clark, B., 2003, EVLA memo 63
``Scaling Relations for Interferometric Post-Processing"

Perley, R.~A., Chandler, C.~J., Butler, B.~J., \& Wrobel, J.~M.,\ 2011, \apjl, 739, L1 

Perley, R.  2004, www.aoc.nrao.edu/evla/geninfo/memoseries/evlamemo64.ps

Perley, R., Napier, P., Jackson, J., et al.\ 2009, IEEE Proceedings, 97, 1448 

Ryle, M. \& Neville, A.C.,1962, MNRAS, 125, 39

Schilizzi, R. T. et al., 2007, SKA memo 100, http://www.skatelescope.org
 ``Preliminary Specifications for the Square Kilometre Array" 

Sault, R.J., Teuben, P.J., \& Wright, M.C.H., 1995,
  in Astronomical Data Analysis Software and Systems IV, ed. R. Shaw, H.E.
  Payne, \& J.J.E.Hayes, ASP Conf. Ser. 77, 433

Thompson, A.~R., Moran, J.~M. \& Swenson, G.~W., 2001, 2nd ed. New York : Wiley, 2001.
``Interferometry and synthesis in radio astronomy"

Welch, J., et al.\ 2009,  IEEE Proceedings, 97, 1438



Wright, M.C.H., 2005, ``Real Time Imaging", SKA memo 60. http://www.skatelescope.org

Wright, M.C.H., et al. 2006, ``SKA Survey Optimization", SKA memo 81. http://www.skatelescope.org

Wright, M.C.H., \& Corder, S., 2008, ``Deconvolving Primary Beam Patterns from SKA Images", SKA memo 103. http://www.skatelescope.org
 
Wootten, A., \& Thompson, A.~R.\ 2009, IEEE Proceedings, 97, 1463

\clearpage

\begin{figure*}[t]
\begin{center}
\includegraphics[scale=0.5]{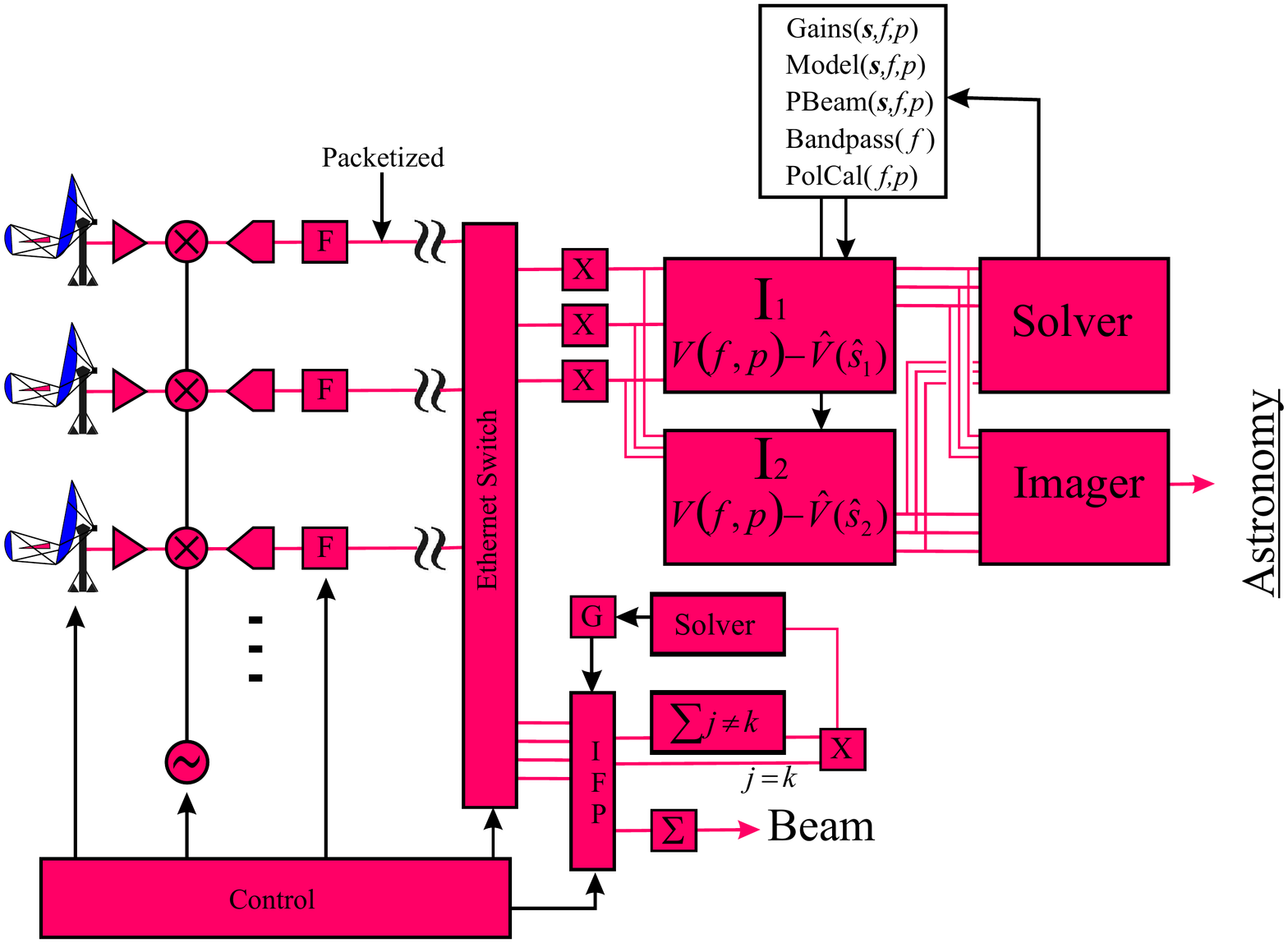}
\caption{\small
   \setlength{\baselineskip}{0.85\baselineskip}
   Data flow from telescopes to images. Signals from each antenna are
converted to baseband (X) and digitized. The sampled bandwidth is divided
into frequency channels using a polyphase filter bank (F). The data are
routed through ethernet switches into cross correlators (X) for each pair
of antennas and frequency channel to measure the correlation properties
of the incident radiation, and into beam formers to form phased array
beams at multiple points in the sky (IFP). The data are calibrated in
the Solver by comparing the measured cross correlations, $V(f,p)$,
with a sky brightness model, $Model(s,f,p)$, to derive instrumental
gains, $Gains(s,f,p)$, primary beams, $PBeam(s,f,p)$, $Bandpass(f)$, and
polarization, $PolCal(f,p)$, calibrations as a functions of position,
$s$, frequency, $f$, and polarization, $p$.  Improvements to the sky
model are made in the Imager from the difference between the calibrated,
measured cross correlations and those derived from the sky model.
  \label{fig:wide-field}
} \end{center} \vskip -0.6in \end{figure*}

\clearpage

\begin{figure*}[t]
\vskip -0.6in
\begin{center}
\includegraphics[scale=0.7]{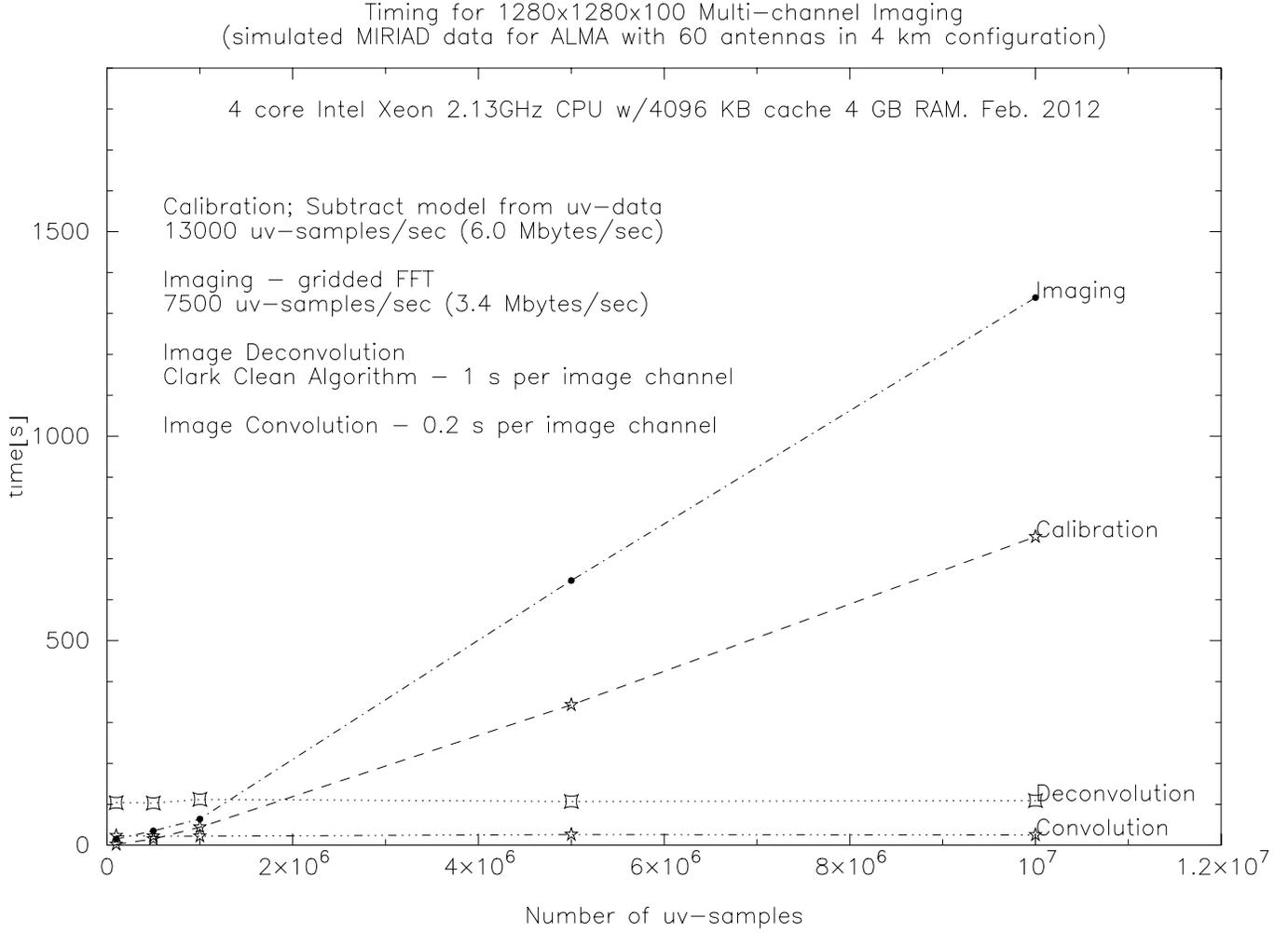}
\caption{\small
   \setlength{\baselineskip}{0.5\baselineskip}
   Timing for 1280x1280x100-channel imaging using simulated MIRIAD data
   for ALMA with 60 antennas in a 4 km configuration.  The dashed line
   shows the time for applying the antenna-based gains and bandpass
   calibrations to the $uv$ data, or for subtracting the sky brightness
   model from the $uv$ data. The imaging step (dash-dotted line) applies
   the weights to the $uv$ data,  convolves the calibrated $uv$ data
   onto a gridded $uv$ plane, and  uses an FFT to make the synthesized
   multi-channel image and  synthesized beam.  Off-line calibration and
   imaging is typically made in several steps.  In a real-time pipeline,
   these steps can be made in sequence on the data stream. The bottom two
   lines show the time for deconvolving the synthesized beam response
   from the 100-channel image using the CLEAN algorithm (dots), and
   convolving by a Gaussian beam (3-dot-dashs).
   \label{fig:timing}
}
\end{center}
\vskip -0.6in
\end{figure*}

\clearpage

\begin{figure*}[t]
\vskip -4.6in
\begin{center}
\includegraphics[scale=0.65]{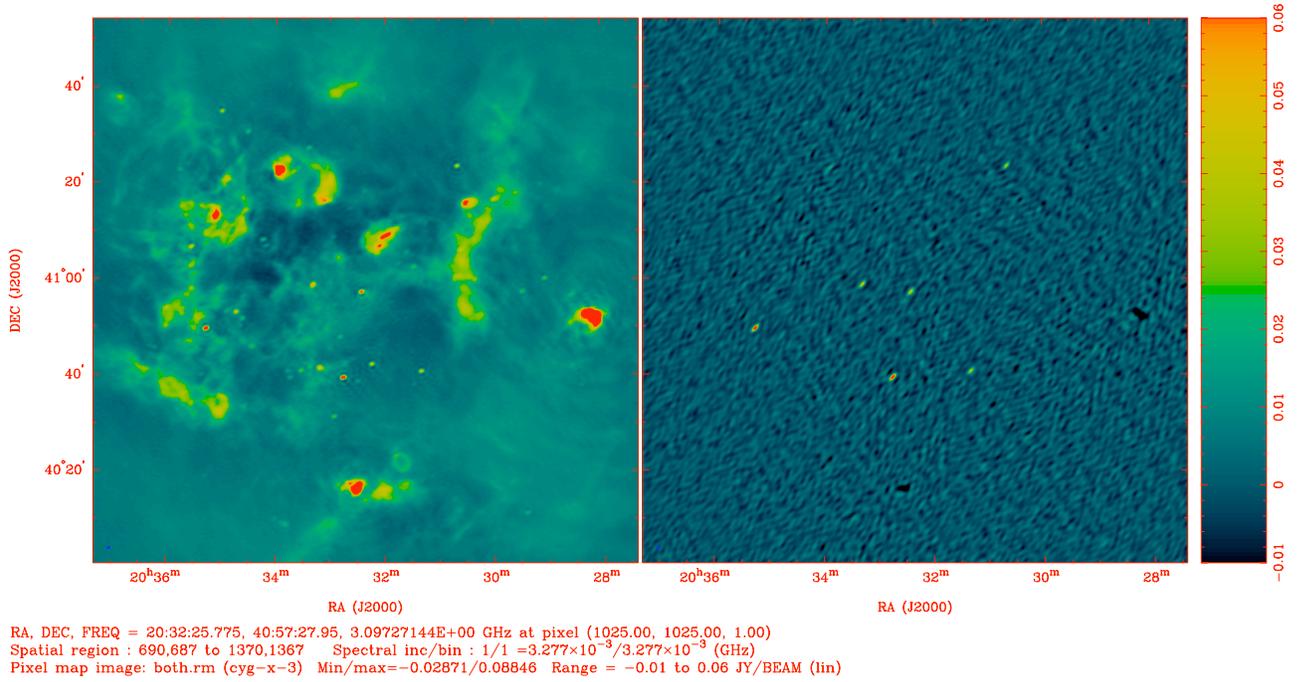}
\vskip -1.0in
\caption{\small
   \setlength{\baselineskip}{0.5\baselineskip}
Left:  Image Cyg X-3 region obtained with the Allen Telescope Array at 3.09 GHz.
Right:  Image of compact, time variable sources after subtracting the complex
 structure. (Peter Williams,  Feb 2011) . Cyg X-3, is a high-mass X-ray
binary system that can increase its brightness by a factor of $\sim$10 in an
hour. Subtracting the large scale structure allows us to get high
time-resolution light curves of time variable sources. The off-line data 
processing took several hours, and limits our ability to image time 
variable sources.
   \label{fig:cygX3}
}
\end{center}
\vskip -0.6in
\end{figure*}
\end{document}